\documentclass[aps,pra,twocolumn]{revtex4-2}
\usepackage{bm}
\usepackage{amsmath}
\usepackage{amsfonts}
\usepackage{amssymb}
\usepackage{graphicx}
\usepackage{color}
\usepackage{xcolor}
\usepackage[colorlinks,linkcolor=blue,anchorcolor=blue,citecolor=blue,urlcolor=blue]{hyperref}
\usepackage{dcolumn}
\usepackage{fancyhdr}
\usepackage[normalem]{ulem}
\usepackage{overpic}
\usepackage{float}

\makeatother
\makeatletter
\newcommand\figurecaption{\def\@captype{figure}\caption}
\newcommand\tablecaption{\def\@captype{table}\caption}
\makeatother
\usepackage[section]{placeins}

\begin{document}

\preprint{APS/123-QED}

\title{Patterning by dynamically unstable spin-orbit-coupled Bose-Einstein condensates}
 
\author{Yunjia Zhai}
\affiliation{Department of Physics, Shanghai University, Shanghai 200444, China}

\author{Yongping Zhang}
\email{yongping11@t.shu.edu.cn}
\affiliation{Department of Physics, Shanghai University, Shanghai 200444, China}

\begin{abstract}
In a two-dimensional atomic Bose-Einstein condensate, we demonstrate Rashba spin-orbit coupling can always introduce dynamical instability into specific zero-quasimomentum states in all parameter regimes.  During the evolution of the zero-quasimomentum states, such spin-orbit-coupling-induced instability can fragment the states and lead to a dynamically patterning process. 
 The features of formed patterns are identified from the symmetries of the Bogoliubov-de Gennes Hamiltonian.  We show that spin-orbit-coupled Bose-Einstein condensates provide an interesting platform for the investigation of pattern formations.
\end{abstract}

\maketitle

\section{Introduction}

Atomic two-component  Bose-Einstein condensates (BECs) are a veritable platform to explore pattern formation.  A key mechanism of patterning is the instability of spatially uniform states against small perturbations~\cite{Cross1993,Gollub1999}.
The instability triggers fast growth of mode-selected perturbations dynamically giving rise to complex spatial structures~\cite{Wang2020}.  When the interactions between components dominate over these of intra-components, a uniform two-component BEC presents phase separation instability~\cite{Law1997,Timmermans1998,Penna2017}.  Experimentally tunable interactions  provide a controllable approach to manipulate the phase separation~\cite{Thalhammer2008}. The uniform two-component BEC with the instability spontaneously breaks to the spatial density patterns of complex alternating domains  due to the immiscible characteristic~\cite{Kasamatsu2004,Ronen2008}. Such pattern formations have been experimentally observed in different two-component setups~\cite{Miesner1999,Papp2008}. A linear coupling between two components causes Rabi oscillations between them and can modify the critical condition for phase separation instability. Coupling-induced pattern dynamics has been observed in experiment~\cite{Nicklas2011}. Furthermore,  the coupling-induced pattern formation has been proposed to test the critical phenomena relating to topological defect formation~\cite{Lee2009,Sabbatini2011,Wu2017}. If the coupling-caused Rabi oscillations are spatially inhomogeneous, a stable moving pattern with antiferromagnetic properties can be generated~\cite{Hamner2013}.

Manipulating two-component BECs with the purpose to introduce other instability mechanisms is often studied for pattern formations.  In spatially segregated two-component BECs, the Rayleigh-Taylor instability ~\cite{Sasaki2009,Gautam2010, Kadokura2012, Xi2018} and the Kelvin-Helmholtz instability~\cite{Kokubo2021} are employed to produce complex patterns along interfaces. The snaking instability of a two-dimensional ring dark soliton leads to different symmetrical patterns with the help of the periodic modulation of the inter-component interactions~\cite{He2019}. The dynamical instability of a linear-coupled two-component polariton condensate is demonstrated to induce complex spatiotemporal patterns with phase dislocations and vortices~\cite{Liew2015}. The periodic modulation of the transverse confinement of a two-component BEC can be used to trigger the Faraday instability~\cite{Staliunas2002,Engels2007, Balaz2012}.  
The formed Faraday patterns, having two different types of  density and spin, are observed in a recent two-component experiment~\cite{Cominotti2022}.  The emergent Faraday patterns in a two-component BEC with parametrically driven dipolar interactions~\cite{Lakomy2012} and in a two-component Fermi-Bose mixtures with driven interactions~\cite{Abdullaev2013} are investigated. An interesting study shows the linear coupling between two components can also lead to the Faraday instability and excite Faraday patterns without the parametric driving~\cite{Chen2019}. Such an idea of Faraday pattern formations is generalized to the Raman-induced spin-orbit coupling instead of the linear coupling in a very recent work~\cite{Zhang2022}. 

On the other hand, a two-component BEC is an ideal platform to study spin-orbit-coupled physics. Spin-orbit coupling can be artificially introduced into BECs using Raman lasers~\cite{Lin2011}. Such a Raman-induced spin-orbit coupling is one-dimensional~\cite{Zhang2016}. While Rashba spin-orbit coupling is two-dimensional. It has been successfully synthesized into two-component BECs~\cite{Wu2016,Sun2018} and degenerate Fermi gases~\cite{Huang2016,Meng2016}.  These experimental advances and abundant spin-orbit-coupled physics revealed by early studies in~\cite{Wang2010,Conjun2011,Ho2011,Hu2012,Yongping2012,Yunli2012} stimulate the wide investigation of  spin-orbit-coupled BECs~\cite{Goldman2014, Goldman2014,Zhai2015,Zhang2014,Li2015,Shizhong2015,Xu2015,Jo2019}. Dynamical instability of a spin-orbit-coupled BEC is a fundamental issue and attracts considerable research interest. Zhu, Zhang and Wu check dynamical instability of  all states in the lower band of a Rashba-coupled BEC and delineate unstable parameter regimes~\cite{Zhu2012}.   Ozawa, Pitaevskii and Stringari examine dynamical instability of  a Raman-induced  spin-orbit-coupled BEC and find that states with the negative effective mass are dynamically unstable~\cite{Ozawa2013}.   Dynamical instability of a spin-orbit-coupled BEC loading into a moving optical lattice is analyzed theoretically~\cite{Zhangyp2013} and experimentally~\cite{Hamner2015}. For various moving velocities of the optical lattice, the instability is experimentally measured by observation of the atom loss in BECs.  Unstable behaviors relate to breakdown of Galilean invariance due to spin-orbit coupling~\cite{Hamner2015}. Mardonov {\it et al.} study instability of a spin-orbit-coupled BEC with attractive interactions and find that spin-orbit coupling can control instability-induced collapse~\cite{Sherman2015}. Much attention has been paid to analyze dynamical instability of a particular state which has a zero quasimomentum~\cite{Wang2010,Bhat2015,Bhuvaneswari2016,Mithun2019,Li2019,Li2017,Singh2020,Tabi2021,Otlaadisa2021,Tabi2022}.  The zero-quasimomentum state is of interest. In the absence of spin-orbit coupling, atoms condensate in this state. Due to the zero quasimomentum, spin-orbit coupling itself has no effect on its existence.  But,  it does make the state dynamically unstable. The spin-orbit-coupling-induced dynamical instability in a zero-quasimomentum state is firstly revealed by Wang {\it et al.}~\cite{Wang2010}.  It has been analyzed in detail for one-dimensional spin-orbit coupling~\cite{Bhat2015,Bhuvaneswari2016,Mithun2019}.  Further relevant studies involve more novel physical environments, such as in the presence of an exotically one-dimensional spin-orbit coupling~\cite{Li2019,Tabi2021,Otlaadisa2021,Tabi2022}, spin-1 spin-orbit coupling~\cite{Li2017}, and Lee-Huang-Yang interactions~\cite{Singh2020}.

In this paper, we reveal that patterns can be formed by the mechanism of  spin-orbit-coupling-induced dynamical instability in a two-dimensional BEC.  We first show that spin-orbit coupling always brings dynamical instability to  specific zero-quasimomentum states in all parameter regimes. This is so-called spin-orbit-coupling-induced instability. 
There are four different zero-quasimomentum states. We classify them basing on whether they carry current or not. Two of them are purely originated from nonlinearity and are unique since spin-orbit coupling is irrelevant to their existence but gives them current.  We are interested in the four states since they are always dynamically unstable in the presence of spin-orbit coupling.  Previous studies~\cite{Wang2010,Bhat2015,Bhuvaneswari2016,Mithun2019,Li2019,Li2017,Singh2020,Tabi2021,Otlaadisa2021,Tabi2022} have already shown the instability of  a no-current-carrying zero-quasimomentum state. We uncover spin-orbit-coupling-induced dynamical instability for these four states by analyzing Bogoliubov-de Gennes (BdG) equations. We then demonstrate that the spin-orbit-coupling-induced instability can trigger  patterning processes for all four states. The current-carrying and no-current-carrying states have different formed patterns.  The geometry of formed patterns is relevant to the symmetry of BdG Hamiltonian. We further reveal that for an anisotropic spin-orbit coupling BdG Hamiltonian of all four states has the same symmetry. So similar patterning processes are found for four states with the anisotropic spin-orbit coupling.  A tunable spin-orbit coupling can be experimentally synthesized into two-component BECs.  Our study demonstrates that a spin-orbit-coupled BEC is an ideal platform for the investigation of pattern formations. 

The paper is organized as follows. In Sec.~\ref{Model}, a theoretical frame to analyze spin-orbit-coupling-induced dynamical instability is provided. It includes Gross-Pitaevskii equations and  Bogoliubov-de Gennes analysis. In
Sec.~\ref{Result}, we prove Rashba coupling can induce dynamical instability to four different zero-quasimomentum states, and show that dynamical instability can trigger patterning for all states. The features of formed patterns relate to the symmetry of BdG Hamiltonian. We furthermore study the case of an anisotropic spin-orbit coupling. Finally, in Sec.~\ref{Conclusion} we summarize our results.

\section{Theoretical model}
\label{Model}

The system is a two-dimensional two-component BEC with spin-orbit coupling.  It is described by the following Gross-Pitaevskii equations~(GPEs),
\begin{equation} 	
\label{GPE}
	i \frac{\partial \Psi}{\partial t}=\frac{p_x^2+p_y^2}{2} \Psi+\left( \delta \sigma_z+ \gamma_{x}p_x\sigma_y-\gamma_{y}p_y\sigma_x\right) \Psi+H_{n}\Psi.
\end{equation}
The spinor is $\Psi(x,y,t)=[\Psi_1(x,y,t), \Psi_2(x,y,t)]^T$ with the first component wave function $\Psi_1$ and the second component $\Psi_2$. $p_x=-i\partial/\partial x$ and $p_y=-i\partial/\partial y$ are the momentum operators along the $x$ and $y$ directions respectively. $(\sigma_x, \sigma_y, \sigma_z)$ are standard Pauli spin-1/2 matrices.  The term $\delta \sigma_z$ represents a Zeeman field along the $z$ direction~\cite{Wu2016}. The two-dimensional spin-orbit coupling is $\gamma_{x}p_x\sigma_y-\gamma_{y}p_y\sigma_x$ with the anisotropic coefficients $\gamma_{x}$ and $\gamma_{y}$. If  $\gamma_{x}=\gamma_{y}$, the coupling becomes Rashba type.  In experiment, the parameters $\delta$, $\gamma_{x}$ and $\gamma_{y}$ are tunable~\cite{Wu2016}.  In the GPEs, the mean-field interactions are described by $H_n$.  
\begin{equation}
H_n=\begin{pmatrix}
g|\Psi_1|^2+g_{12}|\Psi_2|^2&0 \\0& g_{12}|\Psi_1|^2+g|\Psi_2|^2
\end{pmatrix},
\end{equation}
with $g$ and $g_{12}$ being the intra-component and inter-component interaction coefficients respectively. They are proportional to the s-wave scattering lengths. The GPEs in Eq.~(\ref{GPE}) are dimensionless and we use the units of length, energy and time as $1/k_0$, $\hbar^2k_0^2/m$ and $m/(\hbar k_0^2) $ respectively,  here $k_0$ is the wave number of the lasers that are employed to generate spin-orbit coupling~\cite{Wu2016}.

Since the system is spatially homogeneous, stationary solutions of the GPEs are plane waves
\begin{equation}
\label{Plane}
	\Psi(x,y,t)=e^{-i\mu t+ik_xx+ik_y y}\begin{pmatrix}
	\psi_1 \\ \psi_2
	\end{pmatrix}.
\end{equation}
Here, $\mu$ is the chemical potential, $k_x$ and $k_y$ are the quasimomenta along the $x$ and $y$ directions. The spin population, which is spatially independent, satisfies $|\psi_1|^2+|\psi_2|^2=1$.  The nonlinear dispersion relation $\mu(k_x,k_y)$ and the wave functions $(\psi_1, \psi_2)$ can be derived after substituting the plane-wave solutions into Eq.~(\ref{GPE}).

Dynamical instability of these plane-wave solutions can be examined from Bogoliubov-de Gennes (BdG) equations.  After adding perturbations into the plane-wave solutions in Eq.~(\ref{Plane}), general wave functions become,
\begin{align}
\label{Perturbation}
\Psi&(x,y,t)=e^{-i\mu t+ik_xx+ik_y y}   \\
&\times \begin{pmatrix}
\psi_1 +U_1e^{ iq_xx+iq_yy-i\omega t}+V_1^*e^{ -iq_xx-iq_yy+i\omega^* t}\\ \psi_2 +U_2e^{ iq_xx+iq_yy-i\omega t}+V_2^*e^{ -iq_xx-iq_yy+i\omega^* t}
\end{pmatrix},\notag
\end{align}
where $U_{1,2}$ and $V_{1,2}$ are the perturbation amplitudes,  $\omega$ is the perturbation energy, and $q_x, q_y$ are the quasimomenta of perturbation along the $x,y$ directions.  Substituting the general wave functions into Eq.~(\ref{GPE}) and keeping linear terms relating to the perturbation amplitudes, we get the following BdG equations,
\begin{align}
\label{BdG}
\omega\left(\begin{array}{l}
U_{1} \\
V_{1} \\
U_{2} \\
V_{2}
\end{array}\right)=\mathcal{H}_{\mathrm{BdG}}\left(\begin{array}{l}
U_{1} \\
V_{1} \\
U_{2} \\
V_{2}
\end{array}\right).
\end{align}
The BdG Hamiltonian in above is,
\begin{equation}
\mathcal{H}_{\mathrm{BdG}}=\begin{pmatrix}
\mathcal{A}[\psi_1,\psi_2]+\delta\sigma_z &  \mathcal{M}[\psi_1,\psi_2]+\mathcal{T}_{\mathrm{soc}}\\
\mathcal{M}[\psi_2,\psi_1] +\mathcal{T}_{\mathrm{soc}}^*&  \mathcal{A}[\psi_2,\psi_1]-\delta\sigma_z
\end{pmatrix},
\end{equation}
 with, 
  \begin{equation}
 \mathcal{A}[\phi_1,\phi_2]= \left(\frac{q_x^2+q_y^2}{2} -\mu +2g|\phi_1|^2+g_{12}|\phi_2|^2
 \right) \sigma_z, \notag
 \end{equation}
 \begin{equation}
 \mathcal{M}[\phi_1,\phi_2]= \begin{pmatrix}
 g_{12}\phi_1\phi_2^*& g_{12}\phi_1\phi_2\\- g_{12}\phi_1^*\phi_2^*& - g_{12}\phi_1^*\phi_2
\end{pmatrix},\notag
 \end{equation}
and 
 \begin{equation}
\mathcal{T}_{\mathrm{soc}}= -i\gamma_xq_x\sigma_z-\gamma_yq_y\mathbf{I}.\notag
\end{equation}
The BdG Hamiltonian is non-Hermitian, which allows for the existence of complex eigenvalues. For a given state in Eq.~(\ref{Plane}), if $\omega$ in BdG equations have complex modes, the state is dynamically unstable. In the presence of complex modes,  perturbations in Eq.~(\ref{Perturbation}) shall grow up exponentially, which destroys the state.

We focus on the states with zero quasimomentum $k_x=k_y=0$.  We reveal that they are dynamically unstable by calculating BdG equations.  The consequence of their dynamical instability is the formation of density patterns. We show pattern formation by evolving GPEs with initial states as zero-quasimomentum states plus a randomly distributed noise.  The time evolution is implemented by the standard split-step Fourier method.   The window of two-dimensional space is chosen as $(x,y)\in[-51.2, 51.2]$ and is discretized into a $256\times 256 $ mesh grid, and the periodic boundary condition is used for time evolution.

\section{Results and analysis}
\label{Result}

The zero-quasimomentum states in plane-wave solutions Eq.~(\ref{Plane}) are of particular interest.  Their existence does not depend on spin-orbit coupling. Substituting the zero-quasimomentum solutions ($k_x=k_y=0$) in Eq.~(\ref{Plane}) into GPEs, we have
\begin{align}
	& \mu \psi_1 = (g|\psi_1|^2+g_{12}|\psi_2|^2+\delta)\psi_1, \notag \\
		& \mu \psi_2 = (g|\psi_2|^2+g_{12}|\psi_1|^2-\delta)\psi_2.
\end{align}
Solving above nonlinear equations together with the normalization condition $|\psi_1|^2+|\psi_2|^2=1$, we get four different zero-quasimomentum states. They are
\begin{equation}
\label{Solution}
\begin{pmatrix} \psi_1 \\ \psi_2 \\ \mu 
\end{pmatrix}= \begin{pmatrix} 1 \\ 0 \\g+\delta
\end{pmatrix}; \begin{pmatrix} 0\\ 1 \\g-\delta
\end{pmatrix};\begin{pmatrix} \pm  \sqrt{\frac{1}{2}- \frac{\delta}{g-g_{12}} } \\\sqrt{\frac{1}{2}+ \frac{\delta}{g-g_{12}} } \\ \frac{1}{2}(g+g_{12})
\end{pmatrix}. 
\end{equation}
It is noted that the four zero-quasimomentum states do not depend on spin-orbit coupling.  Therefore, their existence is irrelevant to the detail form of spin-orbit coupling. They can also exist if spin-orbit coupling is one dimensional.  The velocity operator of GPEs is calculated as~\cite{Mardonov2015}
\begin{equation}
\hat{\mathbf{v}}=(p_x+\gamma_x\sigma_y)\hat{\mathbf{e}}_x+(p_y-\gamma_y\sigma_x)\hat{\mathbf{e}}_y.
\end{equation}
Current is defined as $\mathbf{J}= Tr\{\hat{\rho} \hat{\mathbf{v}}    \}   $ with pure state density operator $\hat{\rho}= |\Psi \rangle  \langle \Psi| $ . The four states are real-valued, so the current becomes
\begin{align}
\mathbf{J}&=Tr\{ \hat{\rho} \hat{\mathbf{v}} \} = \langle \Psi| \hat{\mathbf{v}} |  \Psi  \rangle \notag \\
&=-\gamma_y \hat{\mathbf{e}}_y \langle \Psi|\sigma_x |  \Psi  \rangle  \\
&= -2\gamma_y \psi_1\psi_2 \hat{\mathbf{e}}_y. \notag
\end{align} 
The current possibly happens along the $y$ direction.  The former two solutions in Eq.~(\ref{Solution}) do not carry current, $\mathbf{J}=0$. In previous studies on dynamical instability in the presence of a one dimensional spin-orbit coupling, only one of these two has been analyzed~\cite{Zhu2012,Ozawa2013,Bhat2015,Bhuvaneswari2016,Mithun2019,Li2019,Li2017,Singh2020,Tabi2021,Otlaadisa2021,Tabi2022}.   The latter two solutions  only exist when $|g-g_{12}|\ge 2\delta$, which indicates that they completely originate from nonlinearity.  They have the same chemical potential and carry opposite currents, $\mathbf{J}=\mp 2\gamma_y \sqrt{\frac{1}{4}- \frac{\delta^2}{(g-g_{2})^2}}\hat{\mathbf{e}}_y$. Even though spin-orbit coupling does not affect the existence of these two nonlinear solutions, it gives them the current which is proportional to $\gamma_y$.

\begin{figure}[t]
	\centering
	\includegraphics[width=0.75\linewidth]{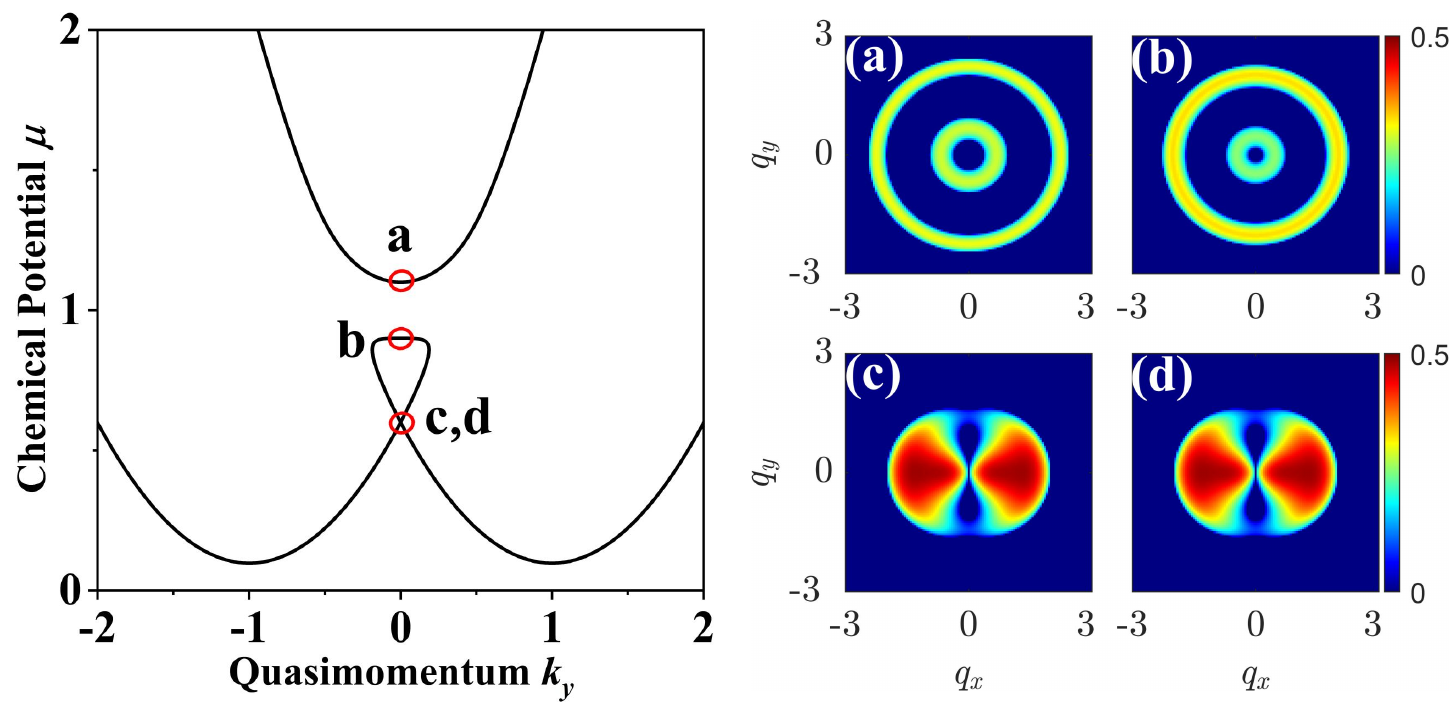}
	\caption{Nonlinear bands and the Rashba spin-orbit-coupling-induced dynamical instability of the zero-quasimomentum states with the parameter regime of $g-g_{12}>2\delta$, and $\gamma_x=\gamma_y=1$. The left panel is the nonlinear bands $\mu (k_x=0,k_y)$ as a function of the quasimomentum $k_y$.   $g=1,g_{12}=0.2$ and  $\delta=0.1$, with these parameters the chemical potential of the current-carrying states labeled by `c' and `d' is lower than the no-current-carrying states labeled by `a' and `b'. The distributions 
		of unstable modes $|\mathrm{Imag}[\omega]|$ (calculated from the BdG equations in Eq.~(\ref{BdG})) for the four zero-quasimomentum states  demonstrated in the perturbation-quasimomentum space $(q_x,q_y)$  in the right panel.}
	\label{Fig1}
\end{figure}

\begin{figure*}[t]
	\centering
	\includegraphics[width=0.75\linewidth]{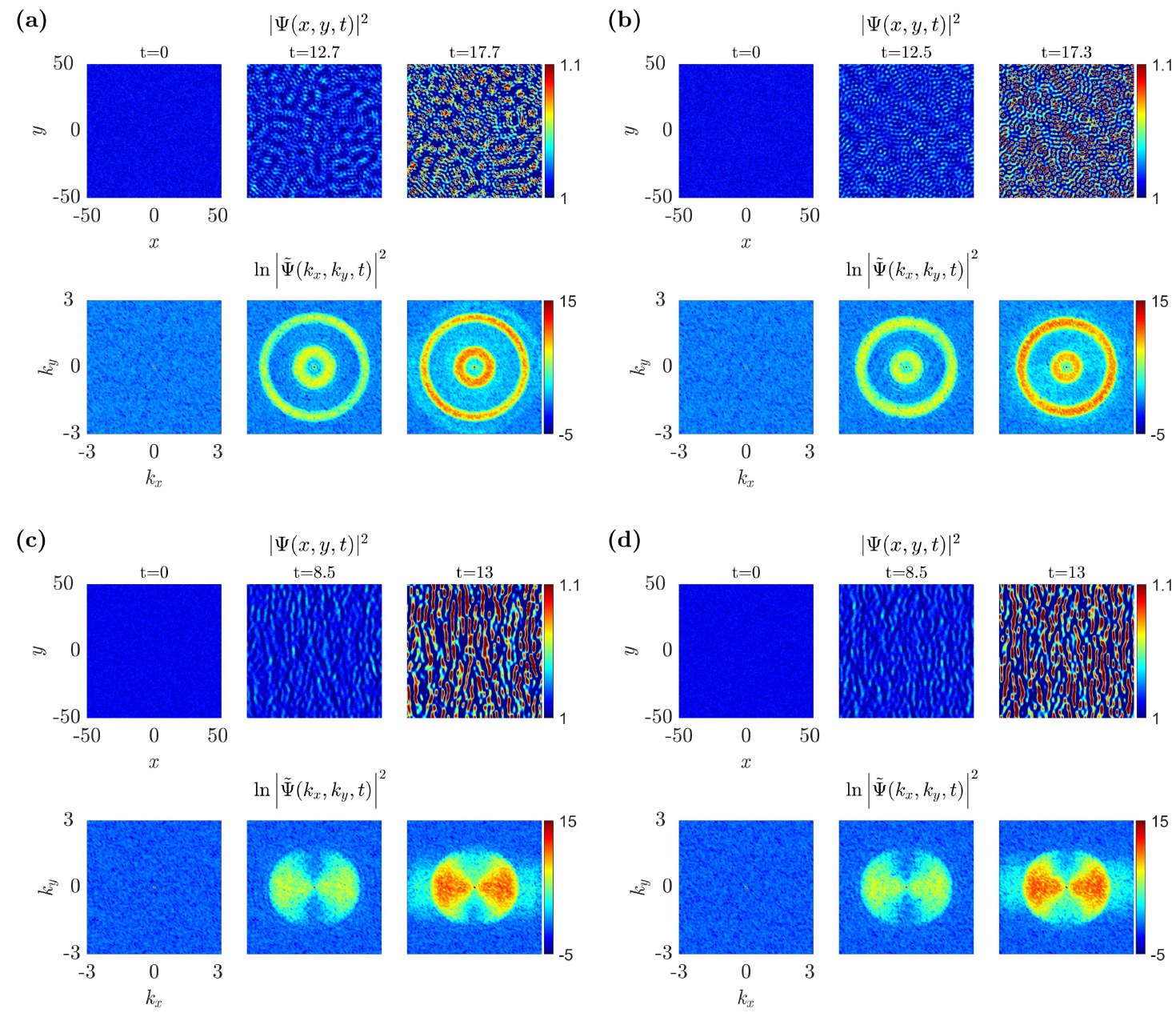}
	\caption{The evolution of the zero-quasimomentum states with $1\%$ uniformly distributed random noise for the Rashba spin-orbit coupling and $g-g_{12}>2\delta$.
		$g=1, g_{12}=0.2 $, $\delta=0.1$ and  $\gamma_x=\gamma_y=1$.  (a) and (b) are for the no-current-carrying states, (c) and (d) are for the current-carrying states. In each plot, the upper panel shows the snapshots of the coordinate-space total density, $|\Psi(x,y,t)|^2=|\Psi_1 (x,y,t)|^2+|\Psi_2 (x,y,t)|^2$. The lower panel demonstrates the snapshot of the logarithm  of the  quasimomentum-space total density,  $\ln |\tilde{\Psi}(k_x,k_y, t)|^{2}=\ln \left\{\left|\tilde{\Psi}_{1}(k_x,k_y, t)\right|^{2}+\left|\tilde{\Psi}_{2}(k_x,k_y, t)\right|^{2}\right\}$.}
	\label{Fig2}
\end{figure*}

\subsection{The Rashba spin-orbit coupling $\gamma_x=\gamma_y$ and  $g-g_{12}>2\delta$}
We  show spin-orbit-coupling-induced dynamical instability for all four states in all parameter regimes.
 We first analyze Rashba coupling, $\gamma_x=\gamma_y$.  The chemical potential of the states depends on nonlinear coefficients and the Zeeman field $\delta$.  When $g-g_{12}>2\delta$, the current-carrying states have a lower chemical potential than that of no-current-carrying states.  In order to show the location of these four states, we calculate full plane-wave solutions with $k_x=0$ and an arbitrary $k_y$. The calculated nonlinear bands $\mu(k_x=0,k_y)$ are demonstrated in the left panel of Fig.~\ref{Fig1} with $g-g_{12}>2\delta$.  There are two bands and a loop structure adhering to the lower band appears.  Nonlinear bands are symmetrical with respect to $k_y=0$. The loop is a pure nonlinear effect,  and its appearance requests $|g-g_{12}|\ge 2\delta$~\cite{Zhangyp2019}.  The no-current-carrying states are two higher states at $k_y=0$, which are labeled by `a' and `b' in the figure.  The current-carrying states locate in the lower parts at $k_y=0$, which are labeled by `c' and `d'. These two states have opposite group velocities, $\partial \mu/\partial k_y \ne 0$,  which is a further indication of current carrying.  
 
 We substitute the four zero-quasimomentum states into BdG equations, and diagonalize the resultant BdG Hamiltonian. Dynamical instability is identified if imaginary parts of $\omega$ are not zero.   In the right panel of Fig.~\ref{Fig1}, we demonstrate the absolute value of  imaginary parts of $\omega$, $|\mathrm{Imag}[\omega]|$, for the four zero-quasimomentum states in the perturbation-quasimomentum $(q_x,q_y)$ plane. The no-current-carrying states [shown in Figs.~\ref{Fig1}(a) and \ref{Fig1}(b)] have a two-ring structure and $|\mathrm{Imag}[\omega]|$ is azimuthally symmetrical. The reason of the azimuthal symmetry can be understood in this way.  We set $q_x=q\cos(\theta)$ and $q_y=q\sin(\theta)$ with $q$ being the magnitude of quasimomentum and $\theta$ being the azimuthal angle.  For the no-current-carrying states, terms relating to the overlap of two components, such as $\psi_1\psi_2$ and $ \psi_1^*\psi_2$, disappear in BdG Hamiltonian, i.e., $\mathcal{M}=0$. A rescaling of perturbation amplitudes,
\begin{equation}
\begin{pmatrix}
U_1\\V_1 \\U_2\\V_2 
\end{pmatrix} \rightarrow  \begin{pmatrix}
e^{-i\frac{\theta}{2}}U_1\\e^{-i\frac{\theta}{2}}V_1 \\e^{i\frac{\theta}{2}}U_2\\e^{-i\frac{3\theta}{2}}V_2 
\end{pmatrix},
\end{equation}
can gauge away the azimuthal angle $\theta$ in BdG equations.   Therefore, $\omega$ does not dependent on $\theta$, giving rise to the azimuthal symmetry in $|\mathrm{Imag}[\omega]|$  shown in Figs.~\ref{Fig1}(a) and \ref{Fig1}(b). Physically, Rashba spin-orbit-coupling  $\gamma_x=\gamma_y$ has a rotational symmetry $e^{i\phi J_z}$ with $\phi$ being an arbitrary angle and $J_z=-i\frac{\partial }{\partial \theta}+\frac{\sigma_z}{2}$. Here $\theta$ is the azimuthal angle defined from  $p_x=p\cos(\theta)$ and $p_y=p\sin(\theta)$ with $p$ being the amplitude.  $[e^{i\phi J_z}, \gamma (p_x\sigma_y-p_y\sigma_x)]=0$. BdG Hamiltonian inherits the symmetry leading to the azimuthal symmetry.

\begin{figure}[b]
	\centering
	\includegraphics[width=0.75\linewidth]{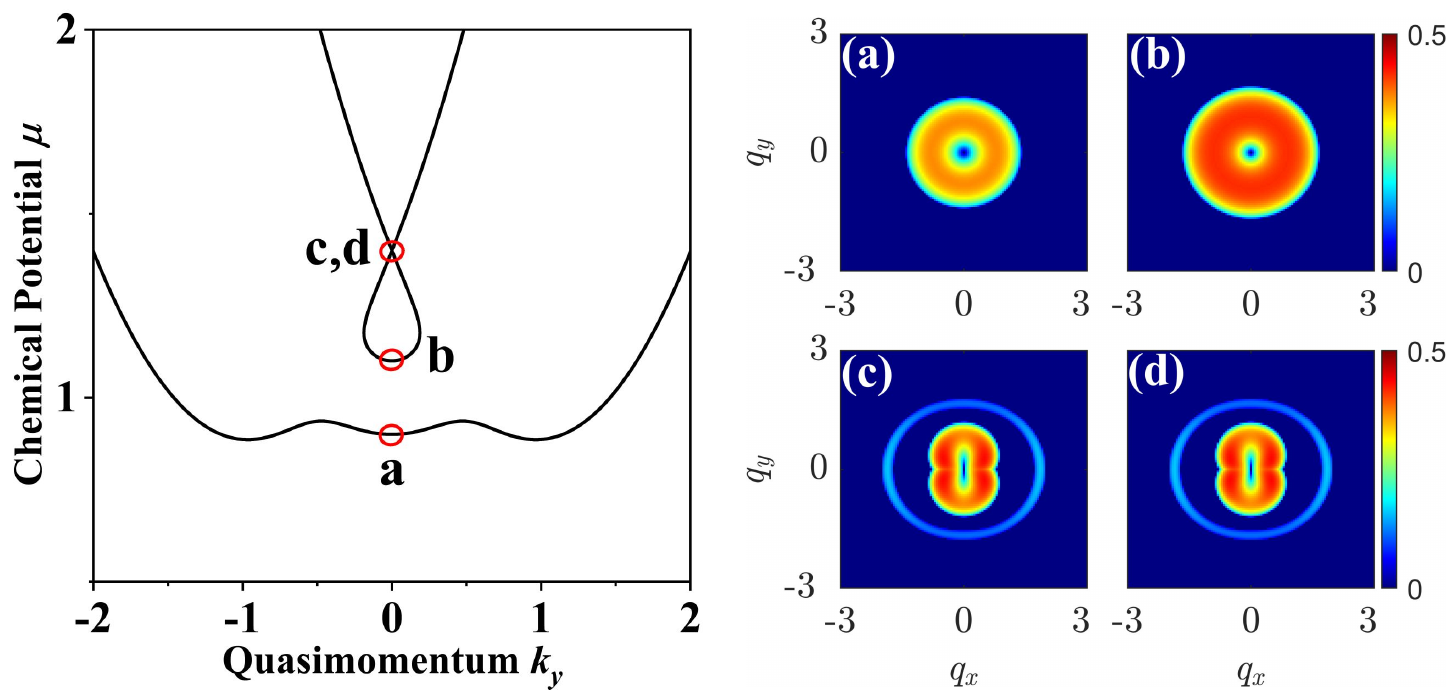}
	\caption{Nonlinear bands and the Rashba spin-orbit-coupling-induced dynamical instability of the zero-quasimomentum states with the parameter regime of $g-g_{12}<-2\delta$, and $\gamma_x=\gamma_y=1$. $g=1,g_{12}=1.8$ and  $\delta=0.1$. The plots show same quantities as in Fig.~\ref{Fig1}. }
	\label{Fig3}
\end{figure}
\begin{figure*}[t]
	\centering
	\includegraphics[width=0.75\linewidth]{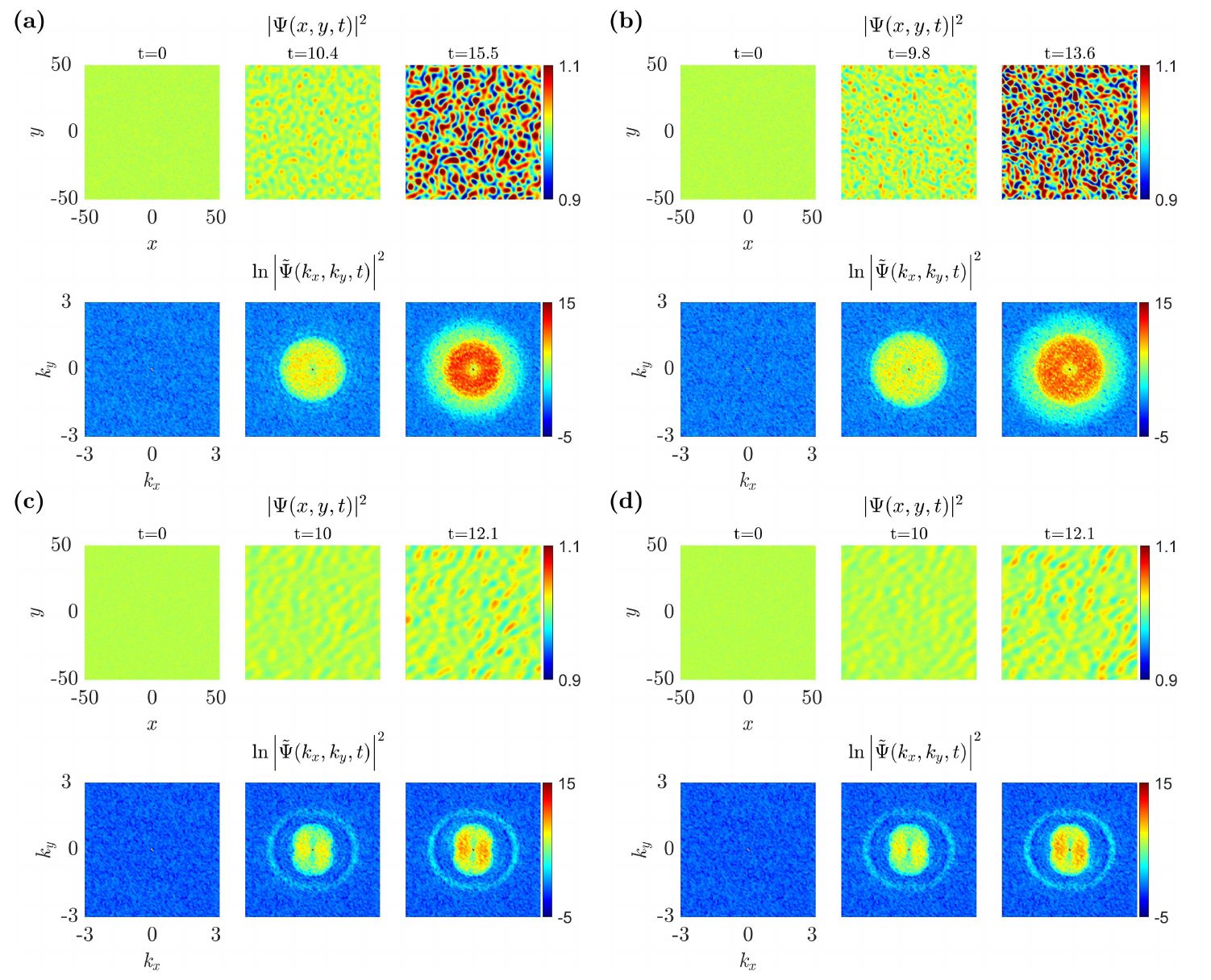}
	\caption{The evolution of the zero-quasimomentum states with $1\%$ uniformly distributed random noise for the Rashba spin-orbit coupling and $g-g_{12}<-2\delta$.
		$g=1, g_{12}=1.8 $, $\delta=0.1$ and  $\gamma_x=\gamma_y=1$.  (a) and (b) are for the no-current-carrying states, (c) and (d) are for the current-carrying states. In each plot, the upper panel shows the snapshots of the coordinate-space total density $|\Psi(x,y,t)|^2$. The lower panel demonstrates the snapshot of the logarithm  of the  quasimomentum-space total density  $\ln |\tilde{\Psi}(k_x,k_y, t)|^{2}$.}
	\label{Fig4}
\end{figure*}

Nevertheless, when terms $\psi_1\psi_2$ in BdG Hamiltonian are non-zero, $ \mathcal{M} \ne 0$ breaks the rotational symmetry, and there does not exist any rescaling to gauge away the azimuthal angle. Consequently, the current-carrying states lose the azimuthal symmetry in $|\mathrm{Imag}[\omega]|$. The calculated results $|\mathrm{Imag}[\omega]|$ are demonstrated in Figs.~\ref{Fig1}(c) and \ref{Fig1}(d) for the current-carrying states. These two states have the same distribution which possesses a $\pi$ rotational symmetry.  The symmetry is because that the wave functions of these two states are real-valued so $\mathcal{H}_{\mathrm{BdG}}^*(q_x,q_y)=\mathcal{H}_{\mathrm{BdG}}(-q_x,q_y)$ is satisfied, which leads to $|\mathrm{Imag}[\omega(q_x,q_y)]|=|\mathrm{Imag}[\omega(-q_x,q_y)]|$.

We have revealed that Rashba spin-orbit coupling can induce dynamical instability into the four zero-quasimomentum states. The unstable modes have interesting distributions in the perturbation-quasimomentum space.  We shall uncover that the dynamical instability can result in the fragmentation of spatially uniform zero-quasimomentum states and leads to pattern formation.  The time evolution of GPEs is implemented by using the initial states as $\Psi_{1,2}(x,y,t=0)=\psi_{1,2} (1+0.01\mathcal{R})$, here $\psi_{1,2}$ are the zero-quasimomentum states and $1\%  $ uniformly distributed random noise $\mathcal{R}$ is added. The role of initial noise is to serve as seeds for boosting unstable perturbation modes. The detailed kind of noise does not affect finally formed patterns but a large amplitude shortens patterning time scale.    In Fig.~\ref{Fig2}, evolution of the coordinate-space  total density, $|\Psi(x,y,t)|^2=|\Psi_1 (x,y,t)|^2+|\Psi_2 (x,y,t)|^2$,  and the logarithm  of the  quasimomentum-space total density,  $\ln |\tilde{\Psi}(k_x,k_y, t)|^{2}=\ln \left\{\left|\tilde{\Psi}_{1}(k_x,k_y, t)\right|^{2}+\left|\tilde{\Psi}_{2}(k_x,k_y, t)\right|^{2}\right\}$, 
are demonstrated in the upper and lower panels of each subplot respectively, here  $\tilde{\Psi}_{1,2}(k_x,k_y, t)=\int dx dy \Psi_{1,2}(x,y,t) e^{-ik_xx-ik_yy}$ are wave functions in the quasimomentum space.  Figs.~\ref{Fig2}(a) and \ref{Fig2}(b) are evolution snapshots for the no-current-carrying states [`a' and `b' in Fig.~\ref{Fig1}]. At $t=0$,  $\ln |\tilde{\Psi}(k_x,k_y, t)|^{2}$ shows a uniformly random distribution since we consider the initial random noise. Around $t=12.7$,  fragmentation of coordinate-space density  $|\Psi(x,y,t)|^2$ leads to a clear pattern. The quasimomentum-space density demonstrates the same structure as unstable modes shown in Figs.~\ref{Fig1}(a) and \ref{Fig1}(b).  This indicates that all the unstable modes are selected from the background noise to grow up.  Around $t=17.7$,  the unstable modes shown in  $\ln |\tilde{\Psi}(k_x,k_y, t)|^{2}$ completely dominate. The pattern of the coordinate-space density is fully established. Since unstable modes have the azimuthal symmetry,  the formed patterns 
are isotropic in the coordinate space.  In Fig.~\ref{Fig2}(b), the occupation of unstable modes in the inner ring is slightly smaller than that in the outer ring. The situation is opposite in  Fig.~\ref{Fig2}(a).  Such difference leads to the distances between patterned density spots in Fig.~\ref{Fig2}(a) are larger  than these in Fig.~\ref{Fig2}(b).  Therefore, the pattern is more dense in Fig.~\ref{Fig2}(b).  Figs.~\ref{Fig2}(c) and \ref{Fig2}(d) demonstrate evolution snapshots for the current-carry states [`c' and `d' in Fig.~\ref{Fig1}]. These two states show a same patterning process. At $t=8.5$,  all unstable modes begin to take shape as shown in  $\ln |\tilde{\Psi}(k_x,k_y, t)|^{2}$.  Meanwhile, the coordinate-space density  is patterning.  At $t=13$, patterns are well established.  The growing modes shown in $\ln |\tilde{\Psi}(k_x,k_y, t)|^{2}$ match with the calculated unstable modes demonstrated in Figs.~\ref{Fig1}(c) and \ref{Fig1}(d). It is noted that the formed patterns for the current-carry states are very different from the no-current-carry states.  Since the unstable modes around  finite $q_x$ and $q_y=0$ dominate [see Figs.~\ref{Fig1}(c) and \ref{Fig1}(d)], the patterns in Figs.~\ref{Fig2}(c) and \ref{Fig2}(d)  become fracted stripes along the $x$ direction, which  also reflects the $\pi$ rotational symmetry of unstable modes in Figs.~\ref{Fig1}(c) and \ref{Fig1}(d).

\begin{figure}[t]
	\includegraphics[width=0.75\linewidth]{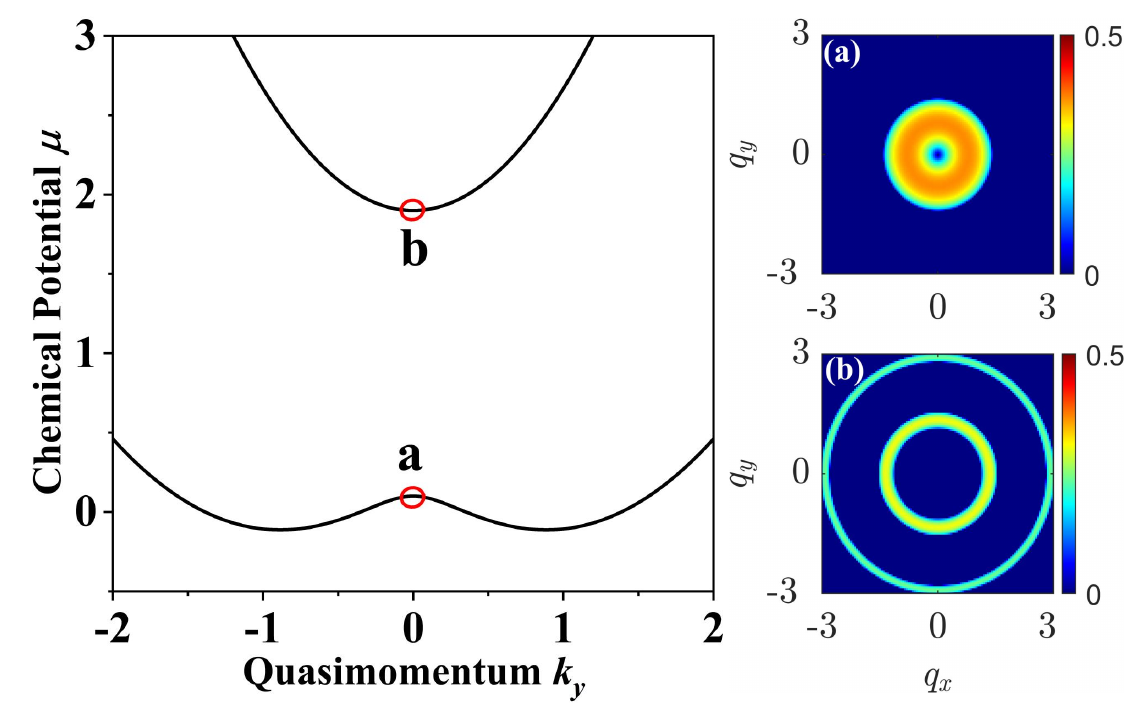}
	\caption{Nonlinear bands and the Rashba spin-orbit-coupling-induced dynamical instability of the zero-quasimomentum states with the parameter regime of $g-g_{12}<2\delta$, and $\gamma_x=\gamma_y=1$. $g=1,g_{12}=0.2$ and  $\delta=0.9$. There are only the no-current-carrying states labeled as `a' and `b'.  The plots show same quantities as in Fig.~\ref{Fig1}. }
	\label{Fig5}
\end{figure}
\begin{figure}[b]
	\centering
	\includegraphics[width=0.75\linewidth]{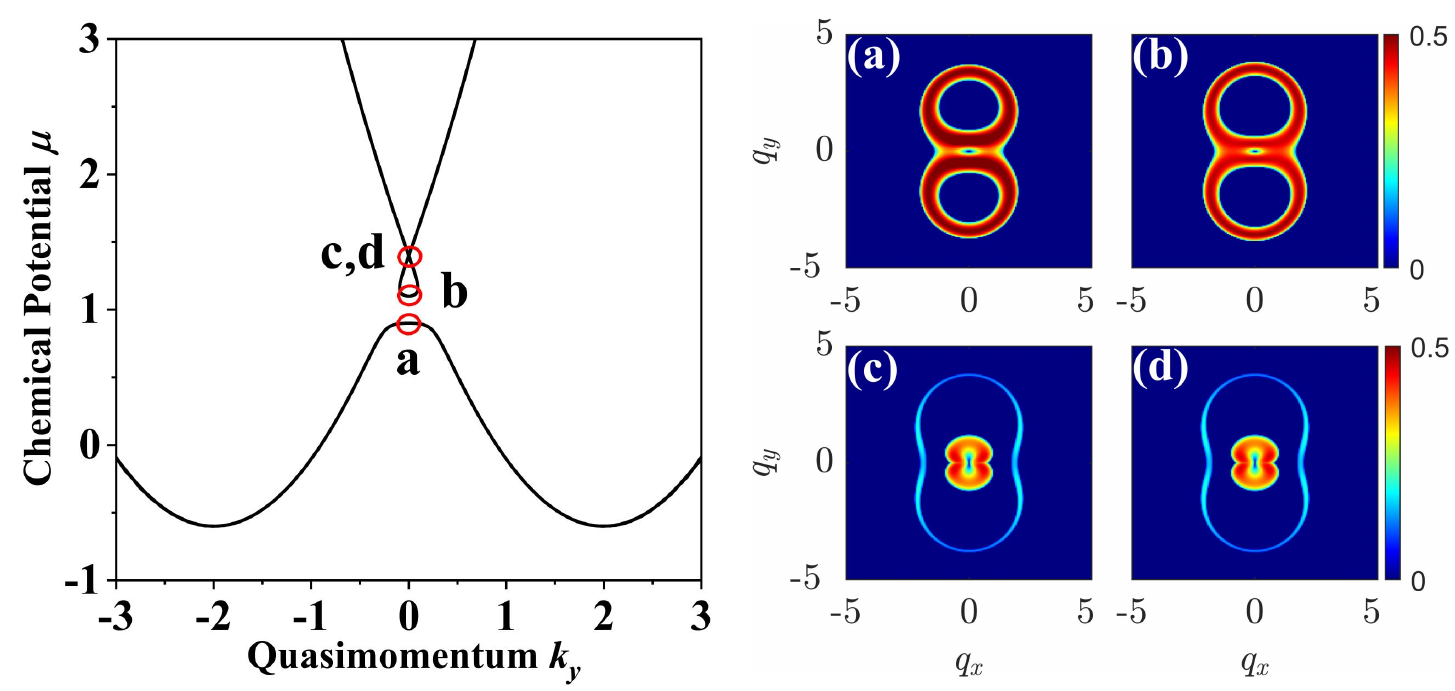}
	\caption{Nonlinear bands and an anisotropic  spin-orbit-coupling-induced dynamical instability of the zero-quasimomentum states with the parameter regime of $g-g_{12}<-2\delta$, and $\gamma_x=1,\gamma_y=2$. $g=1,g_{12}=1.8$ and  $\delta=0.1$. The plots show same quantities as in Fig.~\ref{Fig1}.}
	\label{Fig6}
\end{figure}

\begin{figure*}[t]
	\includegraphics[width=0.75\linewidth]{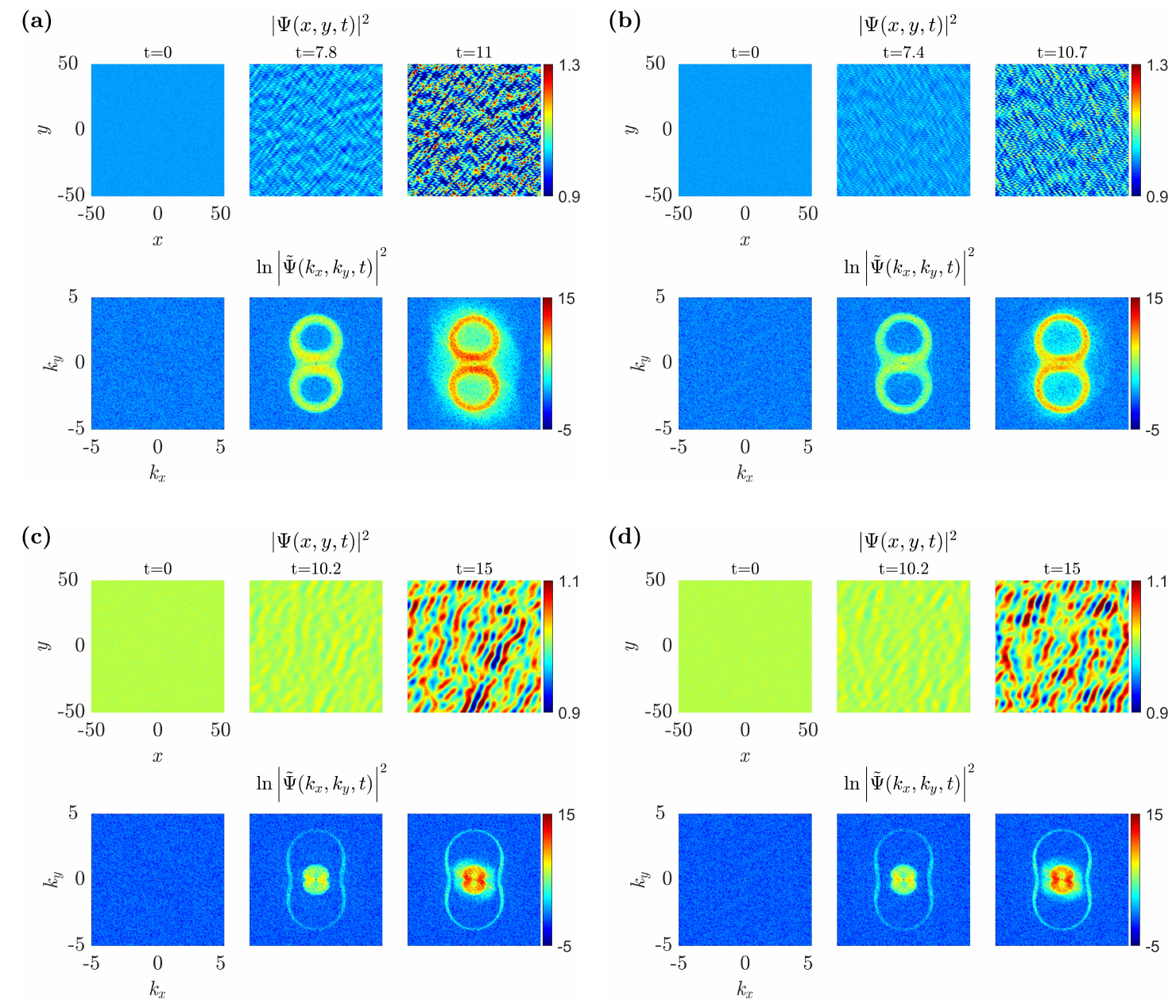}
	\caption{The evolution of the zero-quasimomentum states with $1\%$ uniformly distributed random noise for the anisotropic spin-orbit coupling and $g-g_{12}<-2\delta$.
		$g=1, g_{12}=1.8 $, $\delta=0.1$ and  $\gamma_x=1,\gamma_y=2$.  (a) and (b) are for the no-current-carrying states, (c) and (d) are for the current-carrying states. In each plot, the upper panel shows the snapshots of the coordinate-space total density $|\Psi(x,y,t)|^2$. The lower panel demonstrates the snapshot of the logarithm  of the  quasimomentum-space total density  $\ln |\tilde{\Psi}(k_x,k_y, t)|^{2}$.}
	\label{Fig7}
\end{figure*}

\subsection{The Rashba spin-orbit coupling $\gamma_x=\gamma_y$ and $g-g_{12}<-2\delta$}

We have shown that Rashba-coupling-induced unstable perturbation modes have different symmetries for the no-current-carry and current-carrying states when $g-g_{12}>2\delta$. These unstable modes can grow up from the noisy background, leading to patterned  structures.  The no-current-carrying states have a different patterned geometry from the current-carrying states.  In this section, we study the parameter regime of  $g-g_{12}<-2\delta$, where chemical potential of the current-carrying states is higher than that of the no-current-carrying states.  Typical nonlinear bands $\mu(k_x=0,k_y)$ are demonstrated in the left panel of Fig.~\ref{Fig3}.  The loop structure adheres to the upper band, therefore the current-carry states labeled as `c' and `d' have a large chemical potential. The unstable modes $|\mathrm{Imag}[\omega]|$ are calculated from BdG equations, and are shown in Fig.~\ref{Fig3}. They keep the azimuthal symmetry for the no-current-carrying states and form a disc with a little hole in the middle [see Figs.~\ref{Fig3}(a) and \ref{Fig3}(b)].  For the current-carrying states, unstable modes still have the $\pi$ rotational symmetry, as shown in Figs.~\ref{Fig3}(c) and \ref{Fig3}(d).  However, the distribution is very different from that in Figs.~\ref{Fig1}(c) and \ref{Fig1}(d) in the previous section.

During the time evolution of these zero-quasimomentum states, the unstable modes shown in Fig.~\ref{Fig3} can be spontaneously  selected to grow up.  Figs.~\ref{Fig4}(a) and \ref{Fig4}(b) describe the evolution of the no-current-carrying states. At $t=10.4$ a density pattern becomes obvious in Fig.~\ref{Fig4}(a), while in Fig.~\ref{Fig4}(b), it is at $t=9.8$. The different time scale is because that magnitudes of the unstable modes are different [see Fig.~\ref{Fig3}(a) and \ref{Fig3}(b)]. For the `b' state, $|\mathrm{Imag}[\omega]|$ have larger magnitudes, which results in a faster growth of the unstable modes. The well established patterns at $t=15.5$ in Fig.~\ref{Fig4}(a) and at $t=13.6$ in Fig.~\ref{Fig4}(b) distribute isotropically.  For the current-carrying states shown in Figs.~\ref{Fig4}(c) and \ref{Fig4}(d), the developed pattern is anisotropic.

\subsection{The Rashba spin-orbit coupling $\gamma_x=\gamma_y$ and $|g-g_{12}|<2\delta$}

When $|g-g_{12}|<2\delta$, the current-carrying states in Eq.~(\ref{Solution}) cannot exist.  The zero-quasimomentum states are just two no-current-carrying solutions. They are labeled as `a' and `b' in the full nonlinear bands in Fig.~\ref{Fig5}.  As shown in the left panel, when $|g-g_{12}|<2\delta$,  $\mu(k_x=0,k_y)$ have two nonlinear bands, however, there is no loop structure. The current-carrying states are relevant to the loop, so they do not exist in Fig.~\ref{Fig5}. Rashba spin-orbit coupling introduces dynamical instability to the no-current-carrying states. The distributions of unstable modes $|\mathrm{Imag}[\omega]|$ in the perturbation-quasimomentum space are demonstrated in Figs.~\ref{Fig5}(a) and \ref{Fig5}(b). For the `a' state, unstable modes take a disc geometry with a small hole in the middle. For the `b' state, they appear as two rings.  The time evolution of these two states supports patterning process. The pattern formed from the `a' state is similar to that shown in Figs.~\ref{Fig4}(a) and \ref{Fig4}(b), and formed from the `b' state is similar to that shown in Figs.~\ref{Fig2}(c) and \ref{Fig2}(d).

\subsection{The anisotropic spin-orbit coupling $\gamma_x\ne \gamma_y$}

An anisotropic spin-orbit coupling $\gamma_x\ne \gamma_y$ loses the rotational symmetry $e^{i\phi J_z}$. It is expected that distributions of unstable modes for an anisotropic spin-orbit coupling do not have the azimuthal symmetry.  We study a typically anisotropic case $\gamma_x=1$ and $\gamma_y=2$ in the presence of four zero-quasimomentum states.  The nonlinear bands $\mu(k_x=0,k_y)$ are demonstrated in the left panel of Fig.~\ref{Fig6}. From the nonlinear bands, we identify locations of the zero-quasimomentum states. Since in this case, $g-g_{12}<-2\delta$, the current-carrying states labeled as `c' and `d' in the plot have a larger chemical potential, and the loop adheres to the upper band.  Distributions of unstable modes for the four states are demonstrated in Figs.~\ref{Fig6}(a)-\ref{Fig6}(d).  The particular outstanding is that the distributions for the no-current-carrying states are not azimuthally symmetrical [see Figs.~\ref{Fig6}(a) and \ref{Fig6}(b)].  They are like a digital number eight. While, the distributions for the current-carrying states are also dominated by a digital number eight [see Figs.~\ref{Fig6}(c) and \ref{Fig6}(d)]. Its size is smaller than that of the no-current-carrying states.  For all four states, distributions of unstable modes always have a $\pi$ rotational symmetry since if the wave functions $(\psi_1,\psi_2)$ are real-valued, $\mathcal{H}_{\mathrm{BdG}}^*(q_x,q_y)=\mathcal{H}_{\mathrm{BdG}}(-q_x,q_y)$ can be satisfied, which gives rise to $|\mathrm{Imag}[\omega(q_x,q_y)]|=|\mathrm{Imag}[\omega(-q_x,q_y)]|$. 

The patterning by these four states is demonstrated in Fig.~\ref{Fig7}. The two no-current-carrying states show a similar patterning process [see Figs.~\ref{Fig7}(a) and \ref{Fig7}(b)]. The two current-carrying states present the exact same process [see Figs.~\ref{Fig7}(c) and \ref{Fig7}(d)].  From the quasimomentum-space distributions in each plot, we can know that the selected unstable modes distribute as a digital number eight for all four states. The differences between the states are the size and magnitudes of the number eight, which gives patterns  different length and time scales.  The length scale for the current-carrying states is larger than that for the no-current-carry states. The patterning in Fig.~\ref{Fig7} reflects the same symmetry of BdG Hamiltonian for the four states.

In above, we show that spin-orbit-coupling-induced dynamical instability always exists in the four zero-quasimomentum states for all parameter regimes and the instability leads to patterning processes. Excellence of the zero-quasimomentum states lies in their easily experimental implementations.  Experiments start from a quasi-two-dimensional BEC. The zero-quasimomentum states can be realized by precisely controlling population of two components. Spin-orbit coupling is then suddenly switched on by shining Raman lasers.   The diabatic quench of spin-orbit coupling does not excite the zero-quasimomentum states, since they are also eigenstates of the spin-orbit-coupled system. After holding the system for a certain period during which dynamical instability excites unstable modes to grow up, a time-of-flight measurement is performed to measure momentum-space distributions, from which the coordinate-space patterns can be revealed.

\section{Conclusion}
\label{Conclusion}

Rashba-coupling-induced dynamical instability has been studied in a two-dimensional BEC. There exist four different zero-quasimomentum states. The two of them carry current, and the other two do not. The appearance of these four states does not depend on spin-orbit coupling. However,  we show that the coupling indeed gives them dynamical instability. From BdG equations, we calculate unstable perturbation modes that make the states dynamically unstable.   The momentum-space distributions of unstable modes have different symmetries for the four states. For the no-current-carrying states, the distribution becomes azimuthally symmetrical. While it takes a $\pi$ rotational symmetry for the current-carrying states.   Dynamical instability triggers fast growth of unstable modes from a noisy background, leading to fragmentation of the four homogeneous  states. We show that the fragmentation is accompanied with a patterning process. Affected by the symmetries of unstable modes, the formed coordinate-space density patterns have different features for the four states. Patterns in the no-current-carrying states are isotropic and they have a specific orientation in the current-carrying states.  An anisotropic spin-orbit coupling can also generate dynamical instability into these four states. Unstable modes for all states have the same symmetry in the momentum space. We reveal that the four states have a similar patterning process but with different length and time scales.

\section{Acknowledges}
This work was supported by National Natural Science Foundation of China with Grants No.11974235 and
11774219.


\end{document}